\documentclass[runningheads,a4paper]{llncs}

\usepackage{graphics}
\usepackage{amsmath}
\usepackage{epsfig}
\usepackage{colordvi}
\usepackage[T1]{fontenc}
\usepackage{helvet}
\usepackage{colortbl}
\usepackage{sectsty}
\usepackage{url}
\usepackage{epic}
\usepackage{changebar}
\usepackage{epsfig}
\usepackage{latexsym}
\usepackage{lscape}
\usepackage{amsfonts}
\usepackage{bsymb}
\usepackage{changebar}
\usepackage{tikz}
\usepackage{color}
\usepackage{pdflscape}
\usepackage{setspace}
\usepackage{subfigure}
\usepackage{multirow}
\usepackage[rightcaption]{sidecap}

\newcommand{\Then}{\mbox{\textbf{then}}}
\newcommand{\When}{\mbox{\textbf{when}}}

\newcommand{\beforesection}{\vspace{-8pt}}
\newcommand{\aftersection}{\vspace{-8pt}}

\begin{document}

\mainmatter  

\title{Semi-Automated Design Space \\ Exploration for Formal Modelling}

\titlerunning{Semi-Automated Design Space Exploration for Formal Modelling}

%
%
\author{Gudmund Grov\inst{1} \and Andrew Ireland\inst{1} \and Maria Teresa Llano\inst{2}  \and Peter Kovacs\inst{1} \\ Simon Colton\inst{2} \and Jeremy Gow\inst{2}}
\authorrunning{Grov, Ireland, Llano, Kovacs, Colton and Gow}

\institute{Heriot-Watt University, School of Mathematical and Computer Sciences\\
\email{\{G.Grov,A.Ireland,PK157\}@hw.ac.uk}\\ 
\and
Goldsmiths College, University of London\\
\email{\{m.llano,s.colton,j.gow\}@gold.ac.uk}}
%
%

\toctitle{EDGE}
\tocauthor{Authors}
\maketitle


\begin{abstract}
Refinement based formal methods allow the modelling of systems through incremental steps via abstraction. Discovering the right levels of abstraction, formulating correct and meaningful invariants, and analysing faulty models are some of the challenges faced when using this technique. Here, we propose {\em Design Space Exploration}, an approach that aims to assist a designer by automatically providing high-level modelling guidance in real-time. More specifically, through the combination of common patterns of modelling with techniques from automated theory formation and automated reasoning, different design alternatives are explored and suitable models that deal with faults are proposed. 

\keywords{Formal Methods, Design, Modelling, Refinement, Abstraction, Event-B, Automated Reasoning, Automated Theory Formation.}
\end{abstract}

\beforesection
\section{Introduction} 
\aftersection
\label{sec:intro}
During the development of software intensive systems, the mathematical rigour of {\em formal methods} brings unique benefits. Specifically, the precision of a
formal notation enables design decisions to be clearly communicated
and formally verified.  However, the use of a formal notation alone is
not sufficient to achieve these benefits. Developing design models at
the ``right'' level of abstraction is a creative process,
requiring significant skill and experience on the part of the
designers. Typically within industrial-scale projects, a design will
be modelled at too concrete a level, with the details obscuring the
clarity of key design decisions, making it
harder to determine if the customer's requirements have been
satisfied. In addition, starting with too concrete a design may
prematurely ``lock'' the design team into a particular solution and
increase the complexity of the associated formal verification task,
i.e. proving properties of the design. Addressing these problems would
significantly leverage the creativity of a designer.

We aim at developing 
a tool that analyses the work of a designer behind the scenes, and automatically suggests design
alternatives for Event-B models \cite{JRA10} -- alternatives which improve the clarity and correctness
of a design. Moreover, a tool that explains for each alternative 
{\em what} issue it is addressing and {\em how} it will effect the design
as a whole. 
The tool will be {\em semi-}automatic in that while the analysis and synthesis outlined above will be
automatic, the designer will remain in full control of the design
process. We believe that we can achieve this goal by combining common
patterns of modelling with techniques from automated theory formation and automated reasoning. This paper takes the first steps towards such tool, where the overall  approach is introduced in \S \ref{sec:dse}, together with the implementations of two sub-components and some initial experiments in \S \ref{sec:experiments}. We introduce our running example together with some background on Event-B and automated theory formation in \S \ref{sec:background}, and  conclude with a discussion of related and future work in \S \ref{sec:discussion}.



\section{Background}
\label{sec:background}
\subsection{Event-B}
\label{sec:EventB}
The approach we propose here addresses the challenge of supporting design abstractions, and is relevant to 
formal methods in general, e.g. Alloy \cite{Jackson2012}, B-method \cite{BBook}, Event-B \cite{JRA10}, Z \cite{Spivey92}, VDM \cite{jones-vdm}. 
However, we believe that focusing on a
single formalism will enable us to achieve greater immediate impact, and we
are therefore targeting Event-B, a formalism for modelling discrete complex systems.
Tool support for Event-B is provided via Rodin \cite{Rodin06}, an Eclipse-based 
platform that strongly promotes tool integration via a plug-in style architecture.  

Central to Event-B is the notion of a {\em model} which represents a sequence
of progressively more detailed design abstractions. The dynamic aspects of a
model are represented via a set of \emph{variables} (the state), constrained
by a set of {\em invariants} (which have to be proven to be correct), and a set
of \emph{events}, which are guarded actions. At a given time, any
\emph{enabled} event (an event where the guard is satisfied) can be triggered.

As a working example, consider the requirements given below of a simplified
protocol for transferring money between bank accounts:
\begin{description}
\item[R1:] the sum of money across all accounts should remain constant; 
\item[R2:] transactions can only be completed if the source account has enough funds; 
\item[R3:] if an amount $m$ is debited from a source account, the 
             target account should be credited by $m$; 
\item[R4:] progress should always be possible (no deadlocks).
\end{description}
A designer might choose to represent the protocol as follows in Event-B:
\begin{small}
\begin{tabbing}
$start(a1, a2, m) \stackrel{def}{=}$ \= $\When \ a1 \notin active$ \\
                                    \> $\Then \ pend:=pend \cup \{((a1,a2),m) \} \ ~||~ active:=active \cup \{ a1 \}$ \\
\vspace*{2mm}
$debit(a1, a2, m) \stackrel{def}{=}$ \= $\When \ ((a1,a2),m) \in pend \wedge bal(a1) \geq m$ \\ 
  \> $\Then \ bal(a1):=bal(a1) - m \ || \ pend:=pend \setminus \{((a1,a2),m) \} \ ||$ \\
  \> $\ \ \ \ \ \ \ trans:=trans \cup \{((a1,a2),m) \}$ \\ 
\vspace*{2mm}
$credit(a1, a2, m) \stackrel{def}{=}$ \= $\When \ ((a1,a2),m) \in trans$ \\
  \> $\Then \  bal(a2):=bal(a2)+m \ || \ trans:=trans \setminus \{ ((a1,a2),m) \} \ ||$ \\
  \> $\ \ \ \ \ \ \ active:=active \setminus \{a1 \}$    
\end{tabbing}
\end{small}
The chosen representation involves three steps, each of which is represented through an event that is parametrised by the names of the source ($a1$) and target ($a2$) accounts, along with the value of money ($m$) associated with the transfer. Step one (event $start$) initiates a transfer by adding the transaction to a {\em pending} set ($pend$), and uses a set ($active$) to ensure that an account can only be the source of one transfer at a time. Note that $||$ denotes parallel execution. The second step (event $debit$) removes the funds from the source account if sufficient funds exist -- $bal$ denotes a function that maps an account to its balance. If successful, the transaction is removed from the {\em pending} set and is added to the {\em transfer} set. The final step (event $credit$) completes the transaction by adding the funds to the target account, as well as updating the $trans$ and $active$ sets accordingly. Finally, requirement \textbf{R1} is formalised as an invariant as follows:
$$
\textrm{I1:}~ \Sigma_{a \in dom(bal)} bal(a) = C
$$
where $C$ is a constant that represents the sum of money across all
accounts.

This design abstraction only represents a starting point
for the modelling process. A designer will next refine their design
ideas through a series of progressively more concrete design
abstractions. This gives leverage over the inherent complexity of the
design process, enabling the designer to incrementally achieve a
customer's requirement. Crucially each refinement step must be
formally proved correct. This process is called {\em
  correctness-by-construction}.

\subsection{Automated Theory Formation and The HR System}
Regarding the exploration of the design space, part of our approach consists of building theories of formal specifications based on example simulations of them. This is achieved through the use of Automated Theory Formation (ATF), a technique that invents concepts to describe and categorise examples from the input domain, makes conjectures which relate the concepts, and seeks proofs and counterexamples to determine the truth of the conjectures. In particular, we use HR \cite{colton:atf_book}, an ATF system that 
forms theories about a domain through an iterative application of general purpose Production Rules (PRs) for concept invention. Each PR works by performing operations on the content of one or two input data tables -- where a data table represents a concept with a set of corresponding examples --
and a set of parametrisations in order to produce a new output data table which represents a new concept. 

\begin{figure}[t]
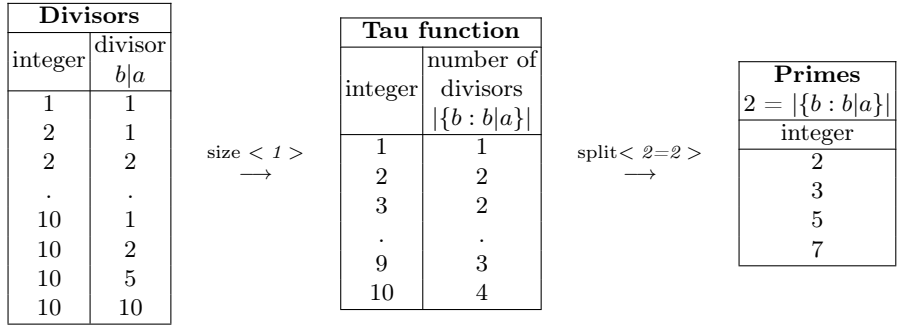

\begin{center}
\begin{footnotesize}
\subfigure{
\begin{tabular}{|c|c|}
\cline{1-2}
\multicolumn{2}{|c|}{\textbf{Divisors}} \\
\cline{1-2}
\multirow{2}{*}{integer} & divisor \\
& $b|a$ \\
\cline{1-2}
1 & 1 \\
2 & 1 \\
2 & 2 \\
. & . \\
10 & 1 \\
10 & 2 \\
10 & 5 \\
10 & 10 \\
\cline{1-2}
\end{tabular}
}
\subfigure{
\begin{scriptsize}
\begin{tabular}{c}
size $<\textit{1}>$ \\
$\longrightarrow$ \\
\end{tabular}
\end{scriptsize}
}
\subfigure{
\begin{tabular}{|c|c|}
\cline{1-2}
\multicolumn{2}{|c|}{\textbf{Tau function}} \\
\cline{1-2}
\multirow{3}{*}{integer} & number of \\
& divisors \\
& $|\{b : b|a\}|$ \\
\cline{1-2}
1 & 1 \\
2 & 2 \\
3 & 2 \\
. & . \\
9 & 3 \\
10 & 4 \\
\cline{1-2}
\end{tabular}
}
\subfigure{
\begin{scriptsize}
\begin{tabular}{c}
split$<\textit{2=2}>$ \\
$\longrightarrow$ \\
\end{tabular}
\end{scriptsize}
}
\subfigure{
\begin{tabular}{|c|}
\cline{1-1}
\textbf{Primes} \\
2 = $|\{b : b|a\}|$ \\
\cline{1-1}
integer \\
\cline{1-1}
2 \\
3 \\
5 \\
7 \\
\cline{1-1}
\end{tabular}
}
\end{footnotesize}
\end{center}
\caption{Steps applied by HR to produce the concept of prime numbers.}
\label{prime}
\end{figure}

To illustrate we show in Figure \ref{prime} the data tables used by HR to produce the concept of prime numbers.
First, HR would take in the concept of divisors ($b|a$ where $b$ is a divisor of $a$), represented by a data table for a subset of integers (partially shown in Figure \ref{prime} for integers from 1 to 10). Then, HR would apply the size PR with the parameterisation $<\textit{1}>$.  This means that the number of tuples 
for each entry in column 1 are counted and recorded in a new data table (the Tau function table). For instance, in the divisors data table 1 appears once in the first column, 2 appears twice, and 10 appears four times. HR then takes in this new concept and applies the split PR with the parameterisation $<\textit{2=2}>$, producing a new data table consisting of those entries in the previous data table whose value in the second column is 2. This is the concept of a prime number.

HR then checks to see whether the data table of the new concept is equivalent to, 
subsumed by, or subsumes another data table, or whether it is empty. Assuming the 
concept of non-square numbers has been formed previously by HR, with a datatable formed by the examples $\{$2,3,5,6,7,8,10$\}$ and the logical construction $\neg(\textit{exists} ~ b.(b|a ~ \& ~ b \ast b=a))$, the data tables of both the concept of prime numbers and the concept of non-square numbers are compared. HR would immediately see that all of its prime numbers are also non-squares, and so conjectures that this
is true for all prime numbers through the following conjecture:  

\begin{center}
\begin{tabular}{ccc}
$\underbrace{2 ~ = ~ |\{b ~ : ~ b|a\}|}$ & $\Rightarrow$ & $\underbrace{\neg(\textit{exists} ~ b.(b|a ~ \& ~ b \ast b=a))}$ \\
{\small prime number} & & {\small non-square number} \\
\end{tabular}
\end{center}

In the context of this paper, ATF is coupled with AR techniques in order to explore the behaviour of the model and guide the search for flaws and alternatives. This is partly motivated by our previous work on \emph{reasoning modelling critics} \cite{jscp11}, a technique where a proof-failure is coupled with a transformation of the model (e.g. add a guard to an event). This technique is limited in that the critics became too syntactic to handle new situations, and we believe our more bottom-up approach of Design Space Exploration will overcome this deficiency. 

We have also used HR to discover invariants of Event-B models \cite{abz12,fac12}. In this work, we guided the search for invariants through proof-failure analysis. This involved prioritising concepts and restricting the PRs to be applied.

\section{Towards Design Space Exploration}
\label{sec:dse}

Key to the style of modelling outlined in \S\ref{sec:EventB} is {\em abstraction} -- the ability to
create a design at the right level of detail; and to ``glue'' it to any abstract model through
a set of gluing invariants.  Trial-and-error is very much part of the expert methodology, where low-level proof
failures are examined, and design alternatives in terms of abstractions are experimented with manually 
(see \cite{Mondex}). Within Design Space Exploration, our goal is to automate much of the low-level grind 
associated with the trial-and-error nature of formal modelling, and provide a designer with {\em high-level} modelling advice in real-time. 

In particular, we aim to generate alternative models at a higher level of \textbf{abstraction} than the original model to deal with a flaw. The intuition is that the flaw is a result of being too concrete. Moreover, within a correct abstraction, the designer has the additional burden of correctly defining the system behaviour
and supplying numerous auxiliary invariants that are required for the formal verification process.
To support this, we will 
suggest \textbf{adaptations} of the initial model at the same level of abstraction. This could be for instance in terms of additional invariants, or even changes to the behaviour of the system.


The unconstrained generation of new models will result in an enormous search space which will be infeasible in practice.
Instead, the approach we are proposing has two phases: 
\begin{enumerate}
\item In an \textbf{analysis} phase the initial model is explored through the application of ATF in order to highlight problematic aspects of the abstraction;  
\item The results of the analysis phase are then used in the model \textbf{generation} phase, where alternative abstractions and adaptations of the 
      model are generated. 
\end{enumerate}
This process may be iterative: the generation may be followed by an analysis phase that 
rules out some ``non-interesting'' alternatives or to identify ``interesting'' features of 
them that require further generations. Below we further explain these phases.


\subsection{Analysis phase} 
A major challenge is to find heuristic techniques that effectively prune the design space so that a designer is presented with a useful set of modelling alternatives. Our heuristic analysis will aim to pin-point both \emph{where} and \emph{what} the problem may be in order to guide the generation phase -- which otherwise will give rise to an explosion of design alternatives -- and to identify the most interesting solutions. Automated Theory Formation (ATF) and Automated Reasoning (AR) will play a heavy role in this analysis.

In order to explore the design space we will allow HR to form unrestricted theories; i.e. without a predefined set of selected PRs. Moreover, we will include event information in order to explore hypotheses related to the events. These two features consist of a significant evolution of our previous work \cite{abz12,fac12}, where the searched space was restricted by enabling specific PRs and the focus was only on the state of the system (not the events). 

The identification of interesting conjectures will consist of different heuristics. Firstly, we will identify conjectures that are associated with failed steps from the simulation trace. This strategy has proven successful as evidenced in \cite{abz12}, and is extended here by including event information. Thus, conjectures of the following forms are sought:
\begin{center}
\begin{tabular}{ccc}
{\bf Variable-based Conjectures} & \hspace{5mm} & {\bf Event-based Conjectures} \\
{\em failure $\Rightarrow$ variable} & & {\em failure $\Rightarrow$ event} \\
{\em failure $\Leftrightarrow$ variable} & & {\em failure $\Leftrightarrow$ event} \\
{\em variable $\Rightarrow$ failure} & & {\em event $\Rightarrow$ failure} \\
{\em variable $\Leftrightarrow$ failure} & & {\em event $\Leftrightarrow$ failure} 
\end{tabular}
\end{center}
which indicate that a variable or an event are associated with failures in the model and therefore should be the focus for the generation phase. In other words, the execution of an event or a change in a variable leads to a failure (event/variable $\Rightarrow$ failure); e.g. an event that always decreases a variable $X$ more than its lower bound limit; or a failure is somehow involved with a variable or an event (failure $\Rightarrow$ event/variable); e.g. the value of a variable $Y$ enables the execution of the event in the previous example. In any of these cases either an adaptation or an abstraction, that involve (all or some of) the identified element(s), should be proposed. This will be further illustrated in \S\ref{sec:experiments}.

Additionally, we will also exploit user given invariants in order to guide the selection process. That is, we seek conjectures that can be identified as correct adaptations of the original invariant; i.e. conjectures of the forms:
\begin{center}
\begin{tabular}{c}
{\bf user\_given\_invariant} $\oplus$ $\lambda$ \\
$\lambda$ $\oplus$ {\bf user\_given\_invariant} \\
$\lambda$ $\oplus$ {\bf user\_given\_invariant} $\oplus$ $\lambda$
\end{tabular}
\end{center}
where $\oplus$ is any correct logical connector, and $\lambda$ can take the form of a single concept or a set of concepts joint themselves by any logical connector. For instance imagine the user given invariant $X \geq C$ which specifies that the value of $X$ must always be greater or equal than some constant $C$, and through the analysis phase the following conjecture is identified: $Y + X \geq C$, where $Y \mapsto \lambda$, $+ \mapsto \oplus$ and $X \geq C \mapsto \textit{user\_given\_invariant}$.

Since we aim at proposing abstractions, we will also search for invariants whose aim is to prove the consistency between the abstract and concrete models; i.e. gluing invariants (a strategy we have already explored in \cite{fac12}). To achieve this we look for conjectures of the forms:
\begin{center}
\begin{tabular}{l}
{\em abstract\_concept} $\Leftrightarrow$ $\lambda$ \hspace{3mm}and\hspace{3mm} $\lambda$ $\Leftrightarrow$ {\em abstract\_concept} \\
\end{tabular}
\end{center}

Finally, we will exploit HR's support for the generation of near conjectures, i.e. conjectures that are true for a percentage threshold of the examples they have. Building upon this functionality, we will explore how this can be tailored to the needs of formal modelling. 
That is, although formal methods are typically based on definite answers, e.g.\ a property is either true or false, we believe that a weaker notion of truth is called for when exploring design alternatives, what we call {\em near-properties}; i.e. properties that are true for most, but not all, behaviours, e.g. {\em ``event $X$ always violates invariant $I$, but it is always re-established by event $Y$''.} Paying attention to such properties can lead to insights of the design and in particular suggest solutions which lie just beyond the fringe of what is currently {\em true} about it. 
\subsection{Generation phase}
The system must be able to `explore' design alternatives also for new and previously unseen scenarios. 
The component that performs the actual generation of new abstractions and adaptations can therefore not be too prescriptive, 
as was the case with our reasoned modelling critics. For his (unpublished) honours dissertation, one of the authors (Kovacs) 
has made the first step towards such a component by implementing a generic framework for model generation as a plug-in of the 
Rodin tool-set \cite{Kovacs15}. The key feature of this plug-in is that it has a layered design: at the bottom is a set of 
low-level but generic `atomic operators' that makes small changes to a model, e.g. `delete variable' and `merge events'.  
These atomic operators can then been combined in order to generate new models. It is up to the system to find the right 
combination of operators. 
Thus, a ``complete'' set of atomic operators would allow the generation of all possible alternative models. 
This gives flexibility to our proposed approach to design space exploration, enabling us to handle new and unforeseen circumstances. 
Figure \ref{fig-model-gen} highlights the atomic operators, conditions and combinators that have been implemented and tested. 
Due to space constraints, we will only summarize this framework and refer the interested reader to \cite{Kovacs15} for further details.
%
\begin{figure}[t]  
\begin{center} 
\subfigure{
{\footnotesize
\begin{tabular}{|l|l|}
\cline{1-2}
\multicolumn{2}{|l|}{\textbf{Operators}} \\
\cline{1-2}
\textbf{Events} & \textbf{Machines} \\
\cline{1-2}
\texttt{deleteArgument}       & \texttt{insertAbstractLayer} \\
\texttt{deleteWitness}        & \texttt{addAbstractLayer} \\
\texttt{deleteGuard}          & \texttt{addSeesContextToAbstractLayer} \\
\texttt{deleteVariable}       & \texttt{moveVariableToAbstractLayer}\\
\texttt{deleteInvariant}      & \texttt{mergeVariable} \\
\texttt{deleteAction}     &  \\
\texttt{negateGuard}      &  \\
\texttt{negateAction}     &  \\
\texttt{mergeEvents}      &  \\
\texttt{combineEvents}    &  \\
\cline{1-2}
\end{tabular}
}
}
\subfigure{
{\footnotesize 
\begin{tabular}{|l|}
\cline{1-1}
\textbf{Conditions} \\
\cline{1-1}
\texttt{hasAbstractLayer} \\
\texttt{hasVariableOfTypePartialFunction} \\
\texttt{hasVariableOfTypePowerset} \\
\texttt{hasAbstractablePartialFunctionVariable} \\
\texttt{hasOneEvent} \\
\cline{1-1}
\end{tabular}
}
}
\subfigure{
{\footnotesize 
\begin{tabular}{|l|}
\cline{1-1}
\textbf{Combinators} \\
\cline{1-1}
\texttt{andApply} \\
\texttt{IfRule} \\
\texttt{IfElseRule} \\
\texttt{RepeatUntilRule} \\
\texttt{WhileRule} \\
\texttt{AccummulatorRule} \\
\cline{1-1}
\end{tabular}
}
}
{\em \begin{quote}
The above operators, conditions and combinators have been implemented and were 
used as the basis for our initial experiments (see \S\ref{sec:experiments}). 
\end{quote} }
\end{center}
\vspace*{-3mm}
\caption{Model generation phase: operators, conditions and combinators.}
\label{fig-model-gen} 
\end{figure} 

Internally, Rodin stores an Event-B model in an abstract syntax tree (AST). Using Scala, this is first translated into an intermediate representation which is essentially a set of triples that describes the relationship between the components. For example, the triple:
\begin{verbatim}
  [Machine(C), refines, Machine(A)]
\end{verbatim}
states that machine \texttt{C} is a refinement of machine \texttt{A}. This creates a more natural and efficient way to implement atomic transformations, as most transformations are small and local.  The operators are then applied to this internal triple representation. For example,
\begin{verbatim}
  Apply deleteVariable On model
\end{verbatim}
will apply the operator that deletes a variable to the Event-B \texttt{model}. Note that applied in this unconstrained way, it will explore all possible alternatives.
That is, one alternative model will be generated for each variable, where each alternative corresponds to the deletion of a single variable. 
We will illustrate more operators in the next section. After the generation, the new alternatives are translated back to Rodin's AST and shown to the user.

Figure \ref{fig-model-gen} lists the operators that have been implemented. Here, operators that are applied to events and machines have been separated. The figure also contains
the implemented \emph{combinators}, which allows us to combine the operators into more complex ones, and \emph{conditions}, which are used to constrain the application of an operator. Examples of these
are given in \S\ref{sec:experiments}. Full details of the operators, coombinators and conditions can be found in \cite{Kovacs15}.

\subsection{Common patterns of modelling} 
As mentioned above, finding the right combination of operators is the key to our approach. 
Common modelling patterns will play a central role in achieving our goal. In comparison with conventional design patterns
(see \cite{paper:Abrial:08b} for examples of design patterns in Event-B), the patterns
we have in mind will be at a much higher level whilst being implemented at a much lower level of 
granularity through the above operator framework. We believe that this will give greater flexibility 
in terms of their application and therefore enable us to provide assistance in situations where
there are no applicable design patterns, or where the application of a
design pattern requires more than just instantiation. 
The analysis will be used to suggest suitable patterns and guidance of how to implement it.
To support this, we have already identified several \emph{refinement patterns} \cite{abz12}
in previous work; however as we cannot refine away flaws, this will be applied in inverse, essentially turning them into 
\emph{abstraction patterns}. Some abstraction patterns have also been identified and represented using the operator framework 
in \cite{Kovacs15}. The experiments in the next section are utilising two patterns:
\begin{enumerate}
\item ``undoing'' bad behaviour by introducing a special \textbf{error} (or exception) \textbf{case}.
\item \textbf{abstracting away} the problem when it can be pinpointed between certain events. This amounts to ``atomising'' sequential events into a single event.
\end{enumerate}

\section{Illustrative examples and initial experiments}
\label{sec:experiments}

\begin{figure}[t]
\includegraphics[width=\textwidth]{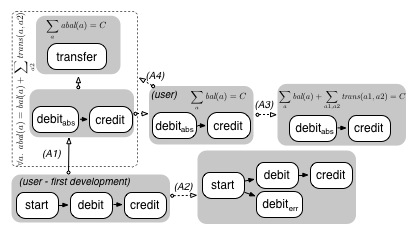} 

{\em Starting from the initial development, abstraction  \textsf{(A1)} and  adaptation \textsf{(A2)}
are suggested to deal with violation of requirement \textsf{R4}. Given that I1 is a near-invariant, a new invariant is suggested in \textsf{(A3)}; or an abstraction \textsf{(A4)} with the required gluing invariant.} 
\caption{A diagrammatic summary of a small design space exploration.}

\label{example-summary}
\end{figure}

In terms of realising our vision we have undertaken experiments at the level of analysing design models as well as mechanising generation. We present these experiments next. The selection of operators and integrations of the phases is currently manual; our ultimate goal is to automate the full development chain.

Consider again the user provided model of a money transfer protocol given in \S\ref{sec:background}. As it stands, the model is flawed since {\bf R4} is violated when all accounts have started a transaction but none of the source accounts have sufficient funds. Moreover, event \emph{debit} violates invariant \textrm{I1} since the amount removed from the source account is not accounted for in the invariant, which makes requirement {\bf R2} broken.  Our aim in such situations will be to offer the designer modelling alternatives which address the flaws. Figure \ref{example-summary} summarises the alternatives generated through our approach, and below we outline how this was achieved.

\paragraph{Abstraction A1}

The first step of the analysis is to provide HR with required model information. This is shown in Figure~\ref{domain} for our running example. The concepts represent the state of the model, with the examples for each of them obtained from simulation traces generated by the ProB simulator \cite{ProB03}. Additionally, ProB checks if the invariants hold at each step of the simulation trace. The concept \textit{good} specifies states for which ProB did not find any invariant violations.

\begin{figure}[t]
\begin{center}
\begin{small}
\hspace{-1cm}
\subfigure{
\begin{tabular}{|l|}
\hline
state(s) \\
\hline
S0 \\
S1 \\
S2 \\
S3 \\
S4 \\
S5 \\
S6 \\
S7 \\
S8 \\
\hline
\end{tabular}
}
\hspace{-3mm}
\subfigure{
\begin{tabular}{|l|}
\hline
integer(i) \\
\hline
0 \\
1 \\
2 \\
... \\
\hline
\hline
event(e) \\
\hline
start \\
debit \\
credit \\
\hline
\end{tabular}
}
\hspace{-3mm}
\subfigure{
\begin{tabular}{|l|}
\hline
account(a) \\
\hline
A1 \\
A2 \\
A3 \\
A4 \\
\hline
\hline
good(x) \\
\hline
S0 \\
S2 \\
S3 \\
S8 \\
\hline
\end{tabular}
}
\hspace{-3mm}
\subfigure{
\begin{tabular}{|l|l|}
\hline
\multicolumn{2}{|c|}{event(s,e)} \\
\hline
S0 & start \\
S1 & debit \\
S2 & credit \\
S3 & start \\
S4 & debit \\
S5 & start \\
S6 & debit \\
S7 & credit \\
S8 & credit \\
\hline
\end{tabular}
}
\hspace{-3mm}
\subfigure{
\begin{tabular}{|l|l|l|}
\hline
\multicolumn{3}{|c|}{bal(s,a,i)} \\
\hline
S0 & A1 & 3 \\
S0 & A2 & 3 \\
S0 & A3 & 3 \\
S0 & A4 & 3 \\
S1 & A1 & 2 \\
S1 & A2 & 3 \\
S1 & A3 & 3 \\
S1 & A4 & 3 \\
... & & \\
\hline
\end{tabular}
}
\hspace{-3mm}
\subfigure{
\begin{tabular}{|l|l|}
\hline
\multicolumn{2}{|c|}{active(s,a)} \\
\hline
S0 & A1 \\
S1 & A1 \\
S3 & A2 \\
S4 & A2 \\
S5 & A2 \\ 
S5 & A3 \\
S6 & A2 \\
S6 & A3 \\
S7 & A3 \\
\hline
\end{tabular}
}
\hspace{-3mm}
\subfigure{
\begin{tabular}{|l|l|l|l|}
\hline
\multicolumn{4}{|c|}{trans(s,a$_{1}$,a$_{2}$,i)} \\
\hline
S1 & A1 & A2 & 1 \\
S4 & A2 & A4 & 1 \\
S5 & A2 & A4 & 1 \\
S6 & A2 & A4 & 1 \\
S6 & A3 & A1 & 3 \\
S7 & A3 & A1 & 3 \\
\hline
\hline
\multicolumn{4}{|c|}{pend(s,a$_{1}$,a$_{2}$,i)} \\
\hline
S0 & A1 & A2 & 1 \\
S3 & A2 & A4 & 1 \\
S5 & A3 & A1 & 3 \\
\hline
\end{tabular}
}
\hspace{-1cm}
\end{small}
\end{center}
\vspace{-20pt}
\caption{Background knowledge provided to HR for the initial analysis phase.}
\label{domain}
\end{figure}
HR is run and in order to prune the output we focus on conjectures related to the states where the failures arise, 
i.e. we search for implications and equivalences that involve the concept $\neg${\em good}. Conjectures 
(\ref{conjActiveone}) are identified, suggesting the generation of bad states associated to event {\em debit} and variable {\em active}: 
\begin{equation}
debit \Rightarrow \neg \textit{good}(a) \qquad 
\neg \textit{good}(a) \Rightarrow \textit{active}(a,b) \label{conjActiveone}
\end{equation}
We can apply the ``abstract away'' pattern to this violation. One implementation of this pattern is to remove the variable that two (sequential) events use
to communicate an intermediate result, and then combine this sequence into an atomic event.  The following Scala code implements such an operator:
\begin{verbatim}
Apply (deleteVariable andApply mergeEvent) On model
\end{verbatim}
We have already discussed \texttt{deleteVariable} above. The \texttt{mergeEvent} operator generates all valid possibilities of combining two events; \texttt{andApply} applies the operators in sequence, such that the second operator is applied to all alternatives generated by the first operator.
Application of the above operator generates 12 alternatives. However, if we include the results from 
the analysis, then we can constrain the operator to the given events and variables using the \texttt{WithActiveElements} filter as follows:
\begin{verbatim}
Apply (deleteVariable andApply mergeEvent) 
   On model WithActiveElements (Event(debit), Variable(pend))
\end{verbatim}
This will generate 2 alternatives, one of them being the desired abstraction:
$$
\begin{array}{rl}
debit_{\textit{abs}}(a1, a2, m) \stackrel{def}{=} &  \ \ \When \ 
a1 \notin active \wedge 
bal(a1) \geq m \\ 
 & \ \ \Then \ active:=active \cup \{ a1 \}  ~||~ 
  bal(a1):=bal(a1) - m \\
  & \ \ \ \ \ \ \ \ \ \ || \ trans:=trans \cup \{((a1,a2),m) \} \\[3pt] 
\end{array}
$$

\paragraph{Adaptation A2}
An alternative analysis is to apply the error-case pattern. Intuitively, this means introducing a new ``error-handling'' event that ``undo'' some previous state changes when the desired path is not applicable. This can be implemented so that it reverses a previous action in cases when an event of the desired
path stays disabled.  This requires new atomic events:
\texttt{combineEvents}, which combines the guards of one event with the actions of another; \texttt{negateGuard}, which negates a guard of an event; and
\texttt{undoActions}, which reverses the actions of an event. An example of the latter is to replace set union by set subtraction. Note that this operator will fail if some action cannot be reversed. Applying these operators in the order specified above (the latter two will only be applied to a combined event) will generate 10 alternatives. However, based on the previous analysis (see A1), we can set event \emph{debit} as active, which reduces the number of alternatives to 7, including the generation of the error-handling event:
$$
\begin{array}{rl}
debit_{err}(a1, a2, m)  \stackrel{def}{=} & \When \ ((a1,a2),m) \in pend
\wedge bal(a1) < m \\
                                    & \Then \ pend:=pend \setminus \{
                                    ((a1,a2),m) \} \ || \\
                                    & \ \ \ \ \ \ \ \ active:=active \setminus \{ a1 \}
                                    \\[3pt]
\end{array}
$$
\noindent which handles the case when the source account does not have enough funds.

\paragraph{Adaptation A3}
Let's assume the user selects  {\bf A1}. Through analysis of this alternative, invariant (I1) is still violated. We run HR on the new model again. The rationale for this is that the user-given invariant (I1) represents a near invariant; therefore, the failed steps would be associated to the elements of the state that make the invariant fail. Through manual inspection we identify the following 2 conjectures:
\begin{equation}
\neg \textit{good}(a) \Leftrightarrow \textit{trans}(a,b,c,d) \qquad 
\neg \textit{good}(a) \Leftrightarrow \textit{active}(a,b) \label{conjActive}
\end{equation}
These conjectures show that the bad states arise when there are transactions currently in progress. In other words, when variables \textit{trans} and \textit{active} are not empty. Based on this we search for conjectures which involve the concepts \textit{trans} and \textit{active} as well as the invariant itself; i.e. \textit{C} = \textit{$\Sigma_{a \in dom(bal)}$ bal(a)}.

We run HR again, this time prioritising the selected concepts. Through manual inspection of the output we identify the following conjecture which is true for all the states of the simulation trace: 
$$
\textit{C}(a,b) \Leftrightarrow (\textit{sum}(\textit{bal},a)=c \wedge \textit{sum}(\textit{trans},a)=d \wedge b = c+d)
$$
This represents an adaptation of invariant \textrm{I1} that addresses the violation by \emph{debit}$_{\textit{abs}}$. Note that this adaptation is achieved by
including the internal state, i.e.  \textit{trans} within the
invariant. The Event-B translation of the invariant, which replaces \textrm{I1}, is:
$$
\textrm{I2:}~\Sigma_{a \in dom(bal)} bal(a) + \Sigma_{(a1,a2) \in
  \textit{dom}(\textit{trans})} \textit{trans}(a1,a2) = C
$$ 

\paragraph{Abstraction A4}

Although correct, invariant \textrm{I2} is not a natural
representation of {\bf R1}, as compared with near-invariant
\textrm{I1}. The designer may wish to explore an alternative
abstraction in which \textrm{I1} is an invariant. Our final 
alternative {\bf A4} represents such an abstraction. 

Based on conjectures 
 (\ref{conjActive}), identified in the output given by HR for alternative  {\bf A1},  we can then re-apply our ``abstract away'' pattern, however its implementation has to be slightly modified as two variables have to be deleted:
\begin{verbatim}
Apply (deleteVariable andApply deleteVariable andApply mergeEvent) 
   On model WithActiveElements (Variable(trans), Variable(active))
\end{verbatim}
Unconstrained, this operator will generate 6 possible alternatives, while the above constrained application, which takes into account the analysis, only generates 2 alternatives, one of them being the desired \textit{transfer} event\footnote{
Technically, the Event-B syntax of the action should be: $abal := abal \triangleleft \hspace{-5pt}- \{a1 \mapsto abal(a1)-m, a2 \mapsto abal(a2)+m \}$}:
$$
\begin{array}{rl}
\mbox{\em{transfer}}(a1,a2,m) \stackrel{def}{=} & \When \ abal(a1) \geq m  \wedge
a1 \neq a2 \\
  & \Then \  abal(a1):=abal(a1)-m \ || \\
        & \ \ \ \ \ \ \ \ abal(a2):=abal(a2)+m
\end{array}$$ 
Finally, in order to prove the consistency between the abstract and concrete models, a gluing invariant is required. Therefore, we enter again in an analysis phase where HR is used to form a theory of the refinement step and search for the invariant. 
By inspecting the output produced by HR, the following conjecture is found:
\begin{align*}
b = sum(trans,a)+sum(bal,a) \Leftrightarrow sum(abal,a)=b
\end{align*}

Although this is not the required invariant, since it compares the total balance instead of the individual balances of each purse, it shows the relation between the abstract variable \textit{abal} and the concrete representation; i.e. variables \textit{bal} and \textit{trans}. Part of our future work will be focused on tailoring HR for the formal methods context so that invariants such as the gluing invariant required in this refinement step can be formed. \\


Instead of proving {\bf R1} directly, Design Space Exploration will
have helped the designer create a more abstract view where the
requirements can be naturally represented. Moreover, we have shown that
this abstraction is respected in the concrete view, i.e. the property holds there too. Note that
\emph{transfer} is also a very natural representation of {\bf R3}, and
by refinement it also holds in {\bf A2}.
\section{Related work, conclusion and future work}
\label{sec:discussion}

Focusing on Event-B, we have introduced our approach to \emph{Design Space Exploration} for formal modelling, which includes a novel integration of automated theory formation, automated reasoning and formal methods. This is supported by an initial implementation with partly automated experiments. We believe this is a novel approach which works 
due to the innovative integration of automated theory formation and automated reasoning with a more traditional formal modelling method (Event-B). This work has been motivated by our previous results on \emph{reasoning modelling critics} \cite{jscp11} and \emph{refinement plans} \cite{abz12}. As the name may suggest, we have taken some inspiration from \emph{automated theory exploration}, which automatically invents theorems and concepts of mathematical theories. We are lifting this to the formal modelling level where we explore alternative models to a given (faulty) user-defined design. A key feature in this is to explore more abstract representations (see e.g. {\bf A1} and {\bf A4}). We are not familiar with any work on automating abstractions within formal modelling, albeit it has been applied to source code (abstract interpretation) and theorem proving (predicate abstraction). Automating refinement, on the other hand, has been studied within several formal methodologies. We have already mentioned our work on refinement plans; other work include: the BART tool for classical B \cite{BART}; the ZRC refinement calculus for Z \cite{ZRC}; and Event-B based tools and techniques as described in e.g. \cite{misc:Iliasov:08,paper:Abrial:08b,misc:Furst:09,abz12}. Contrary to this work, which focuses on automating the refinement to a more concrete step, our proposed approach focus is on automating patterns for abstractions and adaptations. Moreover, as far as we are aware, and different from our work in \cite{fac12},  automated theory formation techniques have not been investigated within the context of refinement style formal modelling. 

Currently, the sub-components of our approach are partly automated, while their integration is manual. HR has to be manually guided and we have to manually inspect its output as well as select and combine the relevant operators to perform the generations. Our goal is to fully automate all parts, and provide users with a set of new (and ideally ordered by perceived relevance) alternatives that they can select. This is where we see the approach as \emph{semi-}automatic: the user will have to make the final decision. In this paper we have provided the first step towards the goal and showed feasibility for the overall approach. However, there is still a long way to go. In \S \ref{sec:dse} and \S \ref{sec:experiments} we have already discussed the next steps for the analysis phase; for the model generation we need to identify a sufficiently small, yet complete, set of atomic operators and combinators to be able to generate all necessary alternatives. It is crucial that these are controlled to avoid generating duplicates. The phases must then be integrated to be able to automate the selection and combination of operators based upon the analysis.

The level of support we aim to provide is very ambitious. If successful, our approach will increase the productivity and accessibility of Event-B, but more importantly, it will provide valuable insights into
how formal methods can be deployed more widely.


\noindent\textbf{Acknowledgments.} 
This work has been supported by EPSRC platform grants \linebreak EP/J001058/1 and EP/N014758/1, and 
FP7 WHIM project 611560. We are grateful for feedback on our approach by Jean-Raymond Abrial. 

\aftersection

\bibliographystyle{abbrv}
\bibliography{ref}

\end{document}